\documentclass[prl,twocolumn,floatfix,superscriptaddress]{revtex4-1}
\usepackage{dcolumn,amsmath}
\usepackage{graphicx}
\usepackage{bm}
\usepackage{hyperref} 

\usepackage{slashbox}
 
\newcommand{\Eeff}{\ensuremath{E_{\rm eff}}}
\newcommand{\eEDM}{{\em e}EDM}
\newcommand{\ecm}{\ensuremath{e {\cdotp} {\rm cm}}}

\newcommand{\de}{d_\mathrm{e}}
\begin{document}
\title{Combined 4-component and relativistic pseudopotential study of ThO for the electron electric dipole moment search}

\author{L.V.\ Skripnikov}\email{leonidos239@gmail.com}
\homepage{http://www.qchem.pnpi.spb.ru}
\affiliation{National Research Centre ``Kurchatov Institute'' B.P. Konstantinov Petersburg Nuclear Physics Institute, Gatchina, Leningrad District 188300, Russia}
\affiliation{Saint Petersburg State University, 7/9 Universitetskaya nab., St. Petersburg, 199034 Russia}
\date{02.10.2016}

\begin{abstract}
A precise theoretical study of the electronic structure of heavy atom diatomic molecules is of key importance to interpret the experiments in the search for violation of time-reversal (T) and spatial-parity (P) symmetries of fundamental interactions it terms of the electron electric dipole moment, \eEDM, and dimensionless constant, $k_{T,P}$, characterizing the strength of the  T,P-odd pseudoscalar$-$scalar electron$-$nucleus neutral current interaction. ACME collaboration has recently obtained and improved limits on these quantities using a beam of ThO molecules in the electronic $H^3\Delta_1$ state [Science 343, 269 (2014)]. We apply the combined direct relativistic 4-component and two-step 
relativistic pseudopotential/restoration  approaches to a benchmark calculation of the effective electric field, \Eeff, parameter of the T,P-odd pseudoscalar$-$scalar interaction, $W_{T,P}$, and hyperfine structure constant in $^3\Delta_1$ state of the ThO molecule. The first two parameters are required to interpret the experimental data in terms of the \eEDM\ and $k_{T,P}$ constant. We have investigated the electron correlation for all of the 98 electrons of ThO simultaneously up to the level of the coupled cluster with single, double and noniterative triple amplitudes, CCSD(T), theory. 
Contributions from iterative triple and noniterative quadruple cluster amplitudes for the valence electrons have been also treated. The obtained values are \Eeff=79.9~GV/cm, $W_{T,P}$=113.1~kHz. The theoretical uncertainty of these values we estimate as about two times smaller than that of our previous study [JCP, 142, 024301 (2015)]. It was found that the correlation of the inner- and outer- core electrons contributes 9\% to the effective electric field. The values of the molecule frame dipole moment of the $^3\Delta_1$ state and the $H^3\Delta_1\to X^1\Sigma^+$ transition energy of ThO calculated within the same methods are in a very good agreement with the experiment.
\end{abstract}

\maketitle

%========================================================================
\section{Introduction.}

The electron can possess a nonzero permanent electric dipole moment 
(electron EDM or \eEDM)
 because of the existence of interactions which violate either the time reversal (T) or spatial parity (P) symmetries, the so-called T,P-odd interactions. Within the standard model (SM) of elementary particles one expects that the \eEDM\ should be 
smaller
than $10^{-38}\ \ecm$ \cite{Khriplovich:11}. However, within most extensions of the SM, the \eEDM\ is expected to have the magnitude within the $10^{-26}-10^{-29}\ \ecm$ range \cite{Commins:98}. Therefore, the 
modern and planned measurements can be considered as test of the SM extensions \cite{Commins:98,Chupp:15}.

It was proposed in the second half of the 20th century 
in Refs.~\cite{Sandars:64, Sandars:65,Labzowsky:78, Sushkov:78, Gorshkov:79, Sushkov:84, Flambaum:85b, Kozlov:87} that the neutral heavy atoms and diatomic molecules
(molecular radicals etc.)
 containing heavy atoms can be efficiently used to search for the \eEDM\ and other T,P-odd effects. Most recently a number of experiments have been performed with Tl \cite{Regan:02}, YbF~\cite{Hudson:11a} and ThO~\cite{ACME:14a} atomic and molecular beams. Some other experiments to search for T,P-odd effects (\eEDM, nuclear quadrupole magnetic moment, nuclear Schiff moment, etc.) are under preparation or 
yet studied theoretically. 
In particular, there are proposals on
 ThO molecule~\cite{FDK14, Skripnikov:14a} as well as on the TaN\cite{FDK14,Skripnikov:15c}, ThF$^+$~\cite{Cornell:13,Skripnikov:15b}, HfF$^+$ \cite{Cossel:12, Cornell:13, Petrov:07a, Fleig:13, Meyer:06a, Skripnikov:08a, Le:13}, PbF \cite{Shafer-Ray:08E, Skripnikov:14c, Petrov:13, Skripnikov:15d}, WC \cite{Lee:13a, Meyer:09a}, RaO \cite{Flambaum:08,Kudashov:13}, RaF \cite{Isaev:12,Kudashov:14}, PtH$^+$ \cite{Meyer:06a, Skripnikov:09}, etc.), TlF~\cite{Skripnikov:09a,Petrov:02,Hinds:80a,Laerdahl:97} molecules and cations. 

One of the most important advantages of using the heavy-atom diatomic molecules with unpaired electrons (and nonzero electron momentum) is the existence of very large effective electric fields, \Eeff, acting on the \eEDM\ \cite{Sushkov:78, Gorshkov:79, Sushkov:84, Flambaum:85b, Kozlov:87,Chubukov:14}. In the mentioned experiments, one measures the interaction energy of the \eEDM\ with the internal effective molecular field.
To extract the value of the \eEDM\ from the experimental energy shift one should know the value of the field. However, the latter cannot be obtained from an experiment and
 can be only obtained theoretically.

The first generation of experiments to search for T,P-odd interactions with a molecular beam of ThO molecules in the metastable  $H^3\Delta_1$ electronic state was performed by ACME collaboration in Ref.~\cite{ACME:14a}. It resulted in a new most rigid limit on the \eEDM: 
$|\de|<9\times 10^{-29}$ 
\ecm\ (90\% confidence).
The experiment has also been interpreted in terms of the T,P-odd pseudoscalar$-$scalar electron$-$nucleus 
neutral current interaction dimensionless constant $k_{T,P}$: $k_{T,P} < 5.9\times 10^{-9}$ (90\% confidence). 
According to the studies performed in Ref.~\cite{Pospelov:14} within the SM, 
this interaction can induce even greater T,P-odd effect simulating the \eEDM.
A new generation 
of experiments 
is
under preparation now and it is expected that in the nearest years ACME collaboration will set a new limit on (or
even 
 measure) the \eEDM\ and $k_{T,P}$ constant by an order of magnitude better \cite{Hess:14,Spaun:14,Petrov:14}.
It involves a considerable enhancement of the experimental technique as well as theoretical investigations of systematic effects.
One of the challenges for the theory is 
calculation of the effective electric field in the molecule containing an actinide atom thorium with a high precision. 
For this one should further develop efficient computational schemes and methods to treat relativistic and high order electron correlation effects.
The results of the ThO experiment~\cite{ACME:14a} have been interpreted using the value of \Eeff\ obtained in Ref. \cite{Skripnikov:13c}. The theoretical uncertainty of this calculation was estimated as 15\%. Later 
a new study of the electron correlation effects was performed and the uncertainty was reduce to 7\% \cite{Skripnikov:15a}.
This uncertainty was determined mainly by 
the approximate version of the two-step approach used (see below). 
Here we propose a new combined two-step and direct 4-component relativistic all-electron approach to minimize the remaining uncertainty.

One should note that there is a number of pure 4-component studies of T,P-odd effects in other diatomic molecules with heavy atoms, e.g. YbF~\cite{Abe:14}, HgH~\cite{Sudip:16}, etc. However, in 
these studies the 
electron correlation effects were treated at the level of the coupled cluster with single and double amplitudes, CCSD theory.
Within the present combined approach we investigate the contribution to \Eeff\ and other considered parameters from high order correlation effects up to the level of the coupled cluster with single, double, triple and perturbative quadruple amplitudes, CCSDT(Q), theory.

The
new results
are compared
 with the previous studies of the ThO molecule \cite{Skripnikov:15a,Fleig:14,FleigComment:16} and
 discrepancies between our \cite{Skripnikov:15a} and other studies \cite{Fleig:14,FleigComment:16}
are discussed.

\subsection{Electronic structure of ThO}

The ThO molecule 
contains
 98 electrons. Below we divide them into three groups: 
(i) the inner-core electrons which correspond to 60 $1s^2-4f^{14}$ electrons of Th and 2 $1s^2$ of oxygen; 
(ii) the outer-core electrons which correspond to $5s^2 5p^6 5d^{10}$ electrons of Th;
(iii) 18 valence electrons which correspond to the rest $6s6p7s6d$ electrons of Th and $2s2p$ electrons of oxygen.
Thus 
one can
 distinguish
: (a)
 the all
(98)
  electron correlation calculations, 
(b) 
 36-electron  correlation calculations (where only the outer-core
and valence
  electrons are treated) and
(c) 
  18-electron calculations (where only the valence electrons are treated).
As $1s^2$ electrons of O 
do not practically contribute to the considered properties they can be included to the outer-core group as it was done in Ref. \cite{Skripnikov:15a} and in the generalized relativistic effective core potential calculations (see below) of the present study.

The electronic state of interest for the \eEDM\ experiment is the first excited $H^3\Delta_1$ electronic state of ThO. In the naive ionic model this state corresponds to the $[\dots] 5s^2 5p^6 5d^{10} 6s^2 6p^6 7s^1 6d^1$ effective electron configuration for Th and $1s^2 2s^2 2p^6$ for O. 

The treatment of correlation contributions to \Eeff\ and other considered properties from the inner-core electrons was not carried out earlier for the ThO molecule and is one of the goals of the present study.

\section{Theoretical details}

The measurable energy shift due to the \eEDM\ is determined by the following
\eEDM\
 Hamiltonian:
\begin{eqnarray}
  H_d=-d_e\sum_j\gamma^0_j\bm{\Sigma}_j \bm{E}_j,
 \label{WdFull}  
\end{eqnarray}
where $j$ is the electron number, $d_e$ is the value of the \eEDM, $\bm{E}$ is the total electric field (due to the nucleus and electrons) acting on an electron, 
$
\bm{\Sigma}=
  \left(\begin{array}{cc}
  \bm{\sigma} & 0 \\
  0 & \bm{\sigma} \\
  \end{array}\right)\ 
$, and $\bm{\sigma}$ are the Pauli matrices.
In the Dirac-Coulomb approximation one can also reduce the
\eEDM\
 Hamiltonian to two forms \cite{Lindroth:89}.
In the first form (``strategy'') one has:
\begin{eqnarray}
  H_d^{I,{\rm eff}}=2d_e\sum_j
  \left(\begin{array}{cc}
  0 & 0 \\
  0 & \bm{\sigma}_j \bm{E}_j \\
  \end{array}\right)\ .
 \label{Wd1}
\end{eqnarray}
In the second form 
one has:
%end ls1
\begin{eqnarray}
  H_d^{II,{\rm eff}}= d_e\sum_j\frac{2i}{e\hbar}c\gamma^0_j\gamma_j^5\bm{p}_j^2,
 \label{Wd2}
\end{eqnarray}
where $\gamma^0$, $\gamma^5$ are the Dirac matrices, 
%$
%\bm{\beta}=
%  \left(\begin{array}{cc}
%  I & 0 \\
%  0 & -I \\
%  \end{array}\right)\ 
%$
 and $\bm{p}$ is the momentum operator for electron.
In the Dirac-Coulomb approximation the expectation values of the 
\eEDM\
Hamiltonians given by Eqs.~(\ref{WdFull}), (\ref{Wd1}) and (\ref{Wd2}) with the exact
(eigen)
 wave function are equal.
An advantage of the second form of the EDM Hamiltonian, Eq.~(\ref{Wd2}), is that it is one-electron operator.
The main contribution to the expectation value of the operator given by Eq.~(\ref{Wd1}) is due to the electric field from the heavy atom nucleus~\cite{Lindroth:89}.
One should note that if the Breit interaction between electrons is considered the EDM Hamiltonians (\ref{Wd1}) and (\ref{Wd2}) are incomplete and the additional two-electron operators should be added to them (see Ref.~\cite{Lindroth:89} for details).

To interpret the results of the molecular ThO (or other diatomics, etc.) experiment in terms of the \eEDM\ one should know a parameter called ``the effective electric field on electron'', \Eeff, which cannot be measured. 
For this one can evaluate an expectation value of the T,P-odd operator $H_d$ (see Eqs.~(\ref{WdFull}),(\ref{Wd1}) and (\ref{Wd2})):
\begin{equation}
\label{matrelem}
W_d = \frac{1}{\Omega}
\langle \Psi|\frac{H_d}{d_e}|\Psi
\rangle,
\end{equation}
where $\Psi$ is the wave function of the considered state of the molecule under consideration and
$\Omega= \langle\Psi|\bm{J}\cdot\bm{n}|\Psi\rangle$,
where $\bm{J}$ is the total electronic momentum, $\bm{n}$ is the unit vector along the molecular axis directed from Th to O in the present case ($\Omega=1$ for the considered $^3\Delta_1$ state of ThO).
In these designations $E_{\rm eff}=W_d|\Omega|$.
 
Besides the interaction given by the operator (\ref{WdFull}) there is another T,P-odd interaction. It is a pseudoscalar$-$scalar electron$-$nucleus interaction with the dimensionless constant $k_{T,P}$. 
The interaction is given by the following operator (see \cite{Hunter:91}):
\begin{eqnarray}
  H_{T,P}=i\frac{G_F}{\sqrt{2}}Zk_{T,P}\sum_j \gamma^0_{j}\gamma^5_{j}n(\textbf{r}_j),
 \label{Htp}
\end{eqnarray}
where $G_F$ is the Fermi-coupling constant and $n(\textbf{r})$ is the nuclear density normalized to unity.
The fundamental constant $k_{T,P}$ can be extracted from the experimental data if one knows the molecular constant $W_{T,P}$ 
that
can be calculated by the following formula:
\begin{equation}
\label{WTP}
W_{T,P} = \frac{1}{\Omega}
\langle \Psi|\frac{H_{T,P}}{k_{T,P}}|\Psi
\rangle.
\end{equation}
Both 
\Eeff\ and $W_{T,P}$ parameters cannot be measured 
neither directly nor indirectly (since \eEDM\ and $k_{T,P}$ are unknown)
%end ls2
and have to be obtained from a molecular electronic structure calculation.
Therefore, only an indirect estimation of the accuracy of these two parameters is possible.

\Eeff\ and $W_{T,P}$ parameters are mainly determined by the behavior of a valence wave function in the region close to the heavy atom nucleus. 
These are so-called
 ``Atoms-In-Compounds'' (AIC) properties or characteristics~\cite{Skripnikov:15b,Titov:14a,Zaitsevskii:16a} (since not only measurable properties but also other effective Hamiltonian parameters 
can be considered here which are not always measurable ). 
In a very good approximation the AIC properties are localized on a heavy atom and do not depend on the bonding
electronic
 density in contrast to some other types of properties \cite{Mayer:07,Sizova:08,Sizova:08b,Sizova:09}.
A very important example of the AIC properties is the magnetic dipole hyperfine structure (HFS) constant.
In contrast to \Eeff\ and $W_{T,P}$ it can be measured. The degree of the agreement of the theoretical and experimental values of the HFS constant can also be considered as an indirect measure of uncertainties of other calculated AIC characteristics.
HFS constant $A_{||}$ can be obtained theoretically by the following matrix element:
\begin{equation}
 \label{Apar}
A_{||}=\frac{\mu_{\rm Th}}{I\Omega}
   \langle
   \Psi|\sum_i\left(\frac{\bm{\alpha}_i\times
\bm{r}_i}{r_i^3}\right)
_z|\Psi
   \rangle, \\
\end{equation}
where $\mu_{\rm Th}$ is the magnetic moment of an isotope of the Th nucleus having the spin $I$.
Up to now there are no experimental data for $A_{||}$ of $^{229}$ThO.

To calculate \Eeff, $W_{T,P}$ and $A_{||}$ we have used a 4-component all-electron as well as two-step technique proposed and developed in Refs.~\cite{Titov:06amin,Skripnikov:15b,Skripnikov:16a}. 
The main idea of the latter method is 
division of the whole molecular calculation into two steps. 
At the first step, one accurately considers the valence part of the molecular wave function within the generalized relativistic effective core potential (GRECP) method \cite{Titov:99, Mosyagin:10a,Mosyagin:16}. The inner-core electrons are excluded from the explicit treatment in this stage. In addition valence wave-functions (spinors) are smoothed in the spatial inner core region of a considered heavy atom.
At the second step, one uses the nonvariational procedure developed in \cite{Titov:06amin,Skripnikov:15b,Skripnikov:16a} to restore the correct 4-component  behavior of the valence wave function in the spatial core region of a heavy atom.
The procedure is based on a proportionality of the valence and
low-lying
 virtual spinors in the inner-core regions of heavy atoms.
Note that the procedure has been recently extended to consider not only the atomic and molecular systems but also three-dimensional periodic structures (crystals) in Ref.~\cite{Skripnikov:15a}.
Below 
the two-step approach is called the ``GRECP/Restoration'' approach.
It has a number of 
advantages~\cite{Titov:99,Titov:06amin,Skripnikov:15b,Skripnikov:16a}. 
One of the features is that one can omit the very time- and resource- consuming stage of two-electron integral transformation that includes small components of molecular bispinors. In practice this stage can be the most time-consuming even in comparison with the correlation treatment. The latter can be rather efficiently parallelized while the parallelization of the former can be more difficult in practice.
Another feature is that one can use 1-component (scalar-relativistic) treatment of the valence electrons. 
Within the approximation it is possible to treat some important corrections, e.g. on a basis set extension or high-order correlation 
effects and analyze saturation of the calculation with respect to the basis set size and level of correlation treatment.
In addition, one can use efficient contracted basis sets rather than uncontracted ones (see below). 
In contrast it is not possible to use contracted basis sets for heavy atoms such as Th in the used 4-component {\sc dirac12} code \cite{DIRAC12}. 
On the other hand, some uncertainties in the GRECP/Restoration approach remain~\cite{Skripnikov:15a}. The main one is that at the present time one cannot use the full version of the GRECP operator \cite{Mosyagin:16} in the available public codes. In the current study 
only the valence (semi-local) part of the GRECP operator
was used.
Besides, the 
nonvariational restoration procedure used here
 is not exactly accurate in practice
\cite{Titov:96}.
  One also cannot treat the correlation of the inner-core electrons in the present formulation of the GRECP/Restoration approach. 
In general, they give a small effect. However, at the level of the current treatment this effect is important and should be considered.

In the present paper we propose to use a combination of the 4-component and GRECP/Restoration approaches to be able 
to take into account the correlation effects from all the electrons of ThO as well as to treat the most important part of the high-order correlation effects.

\section{COMPUTATIONAL DETAILS AND DEFINITIONS}

\subsection{Basis sets}

The following basis sets were used for molecular calculations:
(i) The \textbf{Amax} basis set that was used for the main calculation.
This basis set corresponds to the uncontracted CVTZ basis set for Th \cite{Dyall:07,Dyall:12}.
It includes 33 $s-$, 29 $p-$, 20 $d-$, 15 $f-$, 5 $g-$ and 1 $h-$ type Gaussians for Th. For oxygen the Amax basis set corresponds to the contracted aug-ccpVQZ basis set \cite{Dunning:89,Kendall:92} with two removed g-type basis functions, i.e., 
the (13,7,4,3)/[6,5,4,3] basis set.
(ii) The \textbf{Amin} basis set was used for the calculation of the inner-core correlation contributions up to the CCSD(T) level of correlation treatment.
This basis set corresponds to the VDZ basis set \cite{Dyall:07,Dyall:12} for Th. It includes 26 $s-$, 23 $p-$, 17 $d-$, 13 $f-$, 2 $g-$ type Gaussians for Th. For oxygen the Amin basis set corresponds to the contracted ccpVDZ basis set \cite{Dunning:89,Kendall:92}.
(iii) The \textbf{Amid} basis set was used for the estimation of the uncertainty of the inner-core correlation contributions calculated within the Amin basis set.
This basis set corresponds to the combination of the CVDZ and CVTZ basis sets \cite{Dyall:07,Dyall:12} for Th: it includes 33 $s-$, 29 $p-$, 20 $d-$, 14 $f-$, 2 $g-$ type Gaussians.  For oxygen, the Amid basis set corresponds to the contracted ccpVTZ basis set \cite{Dunning:89,Kendall:92}.
(iv) The \textbf{AmaxExt} basis set was used for the calculation of the basis set extension correction. 
This basis set corresponds to the combination of the CVTZ and CVQZ basis sets \cite{Dyall:07,Dyall:12} for Th: it includes 37 $s-$, 34 $p-$, 26 $d-$, 23 $f-$, 5 $g-$ and 1 $h-$ type Gaussians.  For oxygen, the AmaxExt is equal to the Amax basis set.
(v) The \textbf{Gmax} and \textbf{GmaxExt2} basis sets were used to consider the basis set correction on high angular momenta and extension of basis set on oxygen correction.
The Gmax basis set includes  22 $s-$, 17 $p-$, 15 $d-$, 14 $f-$, 5 $g-$ and 1 $h-$ type partly contracted Gaussians on Th. 
The $g-$
 and $h-$ basis functions are the same as in the Amax basis set.
For oxygen, the Gmax basis set is equal to the Amax basis set.
The GmaxExt2 basis set includes  additional $g-$, $h-$ and $i-$ type Gaussians on Th and corresponds to the aug-ccpV5Z basis set \cite{Dunning:89,Kendall:92} with one removed h-type basis functions on oxygen.
I.e., for O it is the (15,9,5,4,3,1)/[7,6,5,4,3,1] basis set.
(vi) The \textbf{Gmin} basis set was used to estimate the performance of different correlation methods within the GRECP/Restoration approach. This basis set includes 5 $s-$, 4 $p-$, 2 $d-$, 1 $f-$ contracted Gaussians for Th (and can be written as (20,20,10,10)/[5,4,2,1]), and 4 $s-$, 2 $p-$ for oxygen, ((10, 10)/[4,2]).
It was obtained by the further reduction of the natural CBas basis set from Ref.~\cite{Skripnikov:15a} and using the basis set optimization procedure developed in Ref. \cite{Mester:15}.
(vii) The \textbf{Gmid} basis set was used to calculate the contribution from the high-order correlation effects.
This basis set is an extension of the natural CBasSO basis set used in Ref.~\cite{Skripnikov:15a}.
This basis set includes 10 $s-$, 8 $p-$, 5 $d-$, 3 $f-$ contracted Gaussians for Th (and can be written as (25,29,50,10)/[10,8,5,3]), and 6 $s-$, 4 $p-$, 2$d-$ for oxygen, ((16, 10, 6)/[6,4,2]).

We have also used the CBas and MBas basis sets from Ref.~\cite{Skripnikov:15a} in the present study for different purposes (see below).
Note that basis sets used in GRECP calculations contain only functions that are necessary to describe an outer-core and valence part of the wave function because in the vicinity of the Th atom the wave function is smoothed due to the GRECP treatment. Therefore, considerable savings are possible.
Table~\ref{TBasises} gives composition of all used basis sets for convenience.

\begin{table}[!h]
\centering
\caption{Composition of the basis sets used. nS, nP, nD, nF, nG, nH and nI are the numbers of $s-$, $p-$, $d-$,
$f-$, $g-$, $h-$ and $i-$ basis functions.}
\label{TBasises}
\begin{tabular}{lllllllllllllll}
\hline\hline
       & Th &    &    &    &    &    &    &~~~  & O  &    &    &      &  & \\ \cline{1-1}
Basis  & nS & nP & nD & nF & nG & nH & nI &  & S & nP & nD & nF & nG & nH \\ \cline{1-1}
\hline
\\
\multicolumn{15}{l}{Basis sets for \textbf{A}ll-electron 4-component calculations:}                     \\
Amin   & 26 & 23 & 17 & 13 & 2  & 0  & 0  &  & 3  & 2  & 1  & 0  &  & \\
Amid   & 33 & 29 & 20 & 14 & 2  & 0  & 0  &  & 4  & 3  & 2  & 1  &  & \\
Amax   & 33 & 29 & 20 & 15 & 5  & 1  & 0  &  & 6  & 5  & 4  & 3  &  & \\
AmaxExt& 37 & 34 & 26 & 23 & 5  & 1  & 0  &  & 6  & 5  & 4  & 3  &  & \\
\\
\multicolumn{15}{l}{Basis sets for \textbf{G}RECP calculations:}                            \\
Gmax     & 22  & 17  & 15  & 14  & 5  & 1  & 0  &  & 6  & 5  & 4  & 3 & 0  & 0 \\
GmaxExt2 & 22  & 17  & 15  & 14  & 10  & 10  & 5  &  & 7  & 6  & 5  & 4 & 3 & 1   \\
\\
Gmin   & 5  & 4  & 2  & 1  & 0  & 0  & 0  &  & 4  & 2  & 0  & 0  &   & \\
%new
Gmid   & 10  & 8  & 5  & 3  & 0  & 0  & 0  &  & 6  & 4  & 2  & 0  &   & \\
%end new
CBas   & 6  & 5  & 3  & 3  & 0  & 0  & 0  &  & 4  & 3  & 0  & 0  &   & \\
%CBasSO & 6  & 8  & 5  & 3  & 0  & 0  & 0  &  & 4  & 3  & 0  & 0  &   & \\
MBas   & 30 & 8  & 10 & 4  & 4  & 1  & 0  &  & 6  & 5  & 4  & 3  &   & \\
\hline\hline
\end{tabular}
\end{table}

\subsection{Calculation parameters}

To calculate the considered properties we combined our GRECP/Restoration approach with the direct 4-component all-electron approach. The latter allowed us to calculate the contributions from the inner-core shells as well as to avoid an uncertainty due to the use of only the valence part of the full GRECP operator.

In 98-electron correlation calculations we set a cutoff equal to 5000 atomic units for energies of virtual one-electron molecular bispinors.
In 18-electron and 36-electron calculations we set a cutoff equal to 50 atomic units.
For the Th nucleus we used the Gaussian nuclear model with the exponential parameter equal to $1.2897067480\cdot10^{8}$.

In GRECP calculations the $1s-4f$ inner-core electrons of Th were excluded from the molecular correlation calculations using the valence (semi-local) version of the GRECP \cite{Mosyagin:10a,Mosyagin:16} operator. No energy cutoff was used.

For the scalar-relativistic (1-component) GRECP/Restoration 
study
the spin-orbitals used were obtained within the restricted open-shell Hartree-Fock (ROHF) method for the $^3\Delta$ state of ThO.
For 2-component (with included spin-orbit effects) GRECP and 4-component Dirac-Coulomb calculations 
 molecular spinors were obtained using the average-of-configuration Hartree-Fock method for the two electrons in the six spinors (three Kramers pairs).
The latter
 correspond to 7s, 6d$_\delta$ of Th with all other electrons restricted to the closed shells.

We used the following experimental equilibrium internuclear distances \cite{Huber:79,Edvinsson:84}: 3.478 a.u.\ for the $X^1\Sigma^+$ state and 3.511 a.u.\ for the $H^3\Delta_1$ state. As it was shown in our paper \cite{Skripnikov:13c} the calculated equilibrium internuclear distances as well as harmonic frequencies are very close to the experimental values \cite{Huber:79,Edvinsson:84}.

4-component Dirac-Coulomb(-Gaunt) Hartree-Fock calculations were performed within the {\sc dirac12} code \cite{DIRAC12}. Scalar-relativistic coupled cluster (with single, double, and non-iterative triple cluster amplitudes) correlation calculations were performed within the {\sc cfour} code \cite{CFOUR,Gauss:91,Gauss:93,Stanton:97}.
All 4-component coupled cluster calculations as well as scalar-relativistic coupled cluster (CC) and configuration interaction (CI) calculations with the treatment of the high-order cluster amplitudes and excitations were performed within the {\sc mrcc} code  \cite{MRCC2013,Kallay:6}.
The nonvariational restoration code developed by us in Refs.~\cite{Skripnikov:16a, Skripnikov:13c, Skripnikov:11a} and interfaced to these program packages was used to restore the 4-component electronic structure near the Th nucleus.
 
In the present work we have developed and applied the code to compute the matrix elements of the $H_d$~(\ref{Wd2}), $H_{T,P}$~(\ref{Htp}) and HFS~(\ref{Apar}) operators over molecular bispinors.

\section{Results and discussion}
In the present study we 
have
applied the combined scheme which includes the all-electron 4-component and GRECP/Restoration approaches to obtain the most precise values of \Eeff, $W_{T,P}$, A$_{||}$ parameters, molecule-frame dipole moment and $H^3\Delta_1\to X^1\Sigma^+$ transition energy. The main contributions as well as the final values of these parameters are given in Table~\ref{TResults}. 
This Table also includes
the results of the previous studies, in particular, the vibrational contributions corresponding to the zero vibrational level of the $^3\Delta_1$ electronic state of ThO \cite{Skripnikov:15a}.
Below we discuss the obtained results.

\begin{table*}[!h]
\centering
\caption{
Calculated values of the $H^3\Delta_1\to X^1\Sigma^+$ transition energy ($T_e$), molecule-frame dipole moment ($d$), effective electric field (\Eeff), parameter of the pseudoscalar-scalar electron-nucleus interaction ($W_{T,P}$) and hyperfine structure constant (A$_{||}$) 
of the $H^3\Delta_1$ state of ThO using coupled-cluster methods compared to the corresponding values from Refs.~\cite{Fleig:14,FleigComment:16,Skripnikov:15a}.
}
\label{TResults}
\begin{tabular}{llrrrrr}
\hline\hline
Reference & Method                  & $T_e$    & $d$         & \Eeff & $W_{T,P}$ & A$_{||}$                                  \\
                      &                      & ($cm^{-1}$) & (D)      & (GV/cm)  & (kHz)        &($\frac{\mu_{\rm Th}}{\mu_{\rm N}}\cdot$MHz) \\
                    
\hline    

\cite{Fleig:14}       &  VTZ/18e-4c-MR(12)-CISD    & {} 5410 & {}  ---   & {} 75.2  & 106.0~$^a$    &  -2976 \\
                      &  VTZ/18e-4c-MR$_{12}^{+T}$-CISD   & {} --- & {}  ---   & {} 75.2  & 107.8     &  -2880 \\
\cite{FleigComment:16}&    + corrections$^b$    & {}  {}    & {}   &     &    \\
\\    
\cite{Skripnikov:15a} &38e-2c-CCSD(T)                      & {} 5403 & {}  4.23   & {} 81.5  & 112& -2949   \\
                      &    + corrections  $^c$               & {}  {}    & {}   &     &    \\
%\hline     
                
%?????    38e-2c-CCSD+basis corr  &  5002         &  4.19            & 82.9   &    113.8        &   -3027 \\    
%?????    38e-2c-CCSD(T)+basis corr  &  5317         &  4.12            & 81.4   &    111.6        &   -2949 \\
\hline
                   
This work           &CVTZ/18e-4c-CCSD    & 4759      & 4.15         & 76.2   & 107.1      &   -3004    \\                     
                    &CVTZ/18e-4c-CCSD(T) & 5070      & 4.10         & 74.4   & 104.5      &   -2959   \\
                    &CVTZ/36e-4c-CCSD    & 5315      & 4.24         & 80.0   & 112.5      & -3098              \\

\hline
This work           &CVTZ/36e-4c-CCSD(T)$^d$ & 5604      & 4.17         & 78.6   & 110.5      & -3026                                      \\
       &Correlation correction$^f$       & 100        &   0.08         & 0.0   & 0.0        & -2 \\

                    &Inner-core contribution$^d$   &  -5    &    0.01    & 3.6      & 5.0        & -111  \\
            
%            &Basis set correction& -274???    & -0.01    & -1.2     & -0.9       & -41          \\                              
       &Basis set correction 1 (S,P,D,F)$^d$ &     -9     & -0.02   & -0.1    & -0.2    & -2    \\ 
       &Basis set correction 2 (G,H,I and Ox.)$^f$ & -268    & -0.01    & -0.6     & -0.8       & -19 \\                         
       &Gaunt correction$^d$    & -94       & -0.03        & -1.5   & -1.4       & 7       \\       
                    
                    &Vibrational contribution$^f$ \cite{Skripnikov:15a}           & ---       & 0.04         & -0.1   & -0.1       & -2                                         \\
                    &           &              &        &            &                                            \\
                    &Final (this work)    & 5327      & 4.24          & 79.9   & 113.1     & -3155 \\

%Experiment          & 5337      & 4.24 \pm 0.1 &        &            &                                           
\\
                    &Experiment          & {} 5321   & {} 4.098(3)\cite{Hess:14} & {} ---& {} --- &                                             \\ 
                    &    & {}  \cite{Huber:79,Edvinsson:84} & {} 4.24{$\pm$}0.1\cite{Vutha:2011} & {} &  &                                             \\ 
    \hline\hline

\end{tabular}
\\
\flushleft
$^a$ Calculated in Ref.~\cite{FleigComment:16}.\\
$^b$ Corrections on the reference spinors, active spinor space size, Gaunt interaction and outer-core correlation (within the MR3-CISD method), see Ref.~\cite{FleigComment:16} for details.\\
$^c$ Corrections on the basis set extension, high-order correlation effects and vibrational contribution, see Ref.~\cite{Skripnikov:15a} for details.\\
$^d$ Calculated within the 4-component approach.\\
$^f$ Calculated within the two-step GRECP/Restoration approach.
\end{table*}

\subsection{High order correlation effects: 4-component and GRECP/Restoration}

For the precision calculation of \Eeff\ and other considered parameters it is important to investigate the contribution from the high-order correlation effects.
The problem of such a treatment is in a very high computational cost of the methods such as the coupled cluster with single, double, triple and noniterative quadruple cluster amplitudes, CCSDT(Q), method. The complexity of this method  increases rapidly with the basis set size. Therefore, the methods of the construction of compact basis sets as well as a practical possibility to use these basis sets are of high importance.
At present it is only possible to use the uncontracted basis sets for Th in 4-component calculations in the used {\sc dirac12} code \cite{DIRAC12}. These basis sets consist of primitive (single) Gaussians. On the other hand, within 1- and 2-component (GRECP) calculations it is possible to use the contracted basis sets where each basis function is a linear combination of a number of primitive Gaussians. Due to this flexibility, one can construct and use more compact basis sets.

In the present paper 
a combined application of the direct 4-component and GRECP/Restoration approaches
in considered.
Table~\ref{HighCorr} presents
the correlation contributions to \Eeff\ calculated within the 2-component GRECP/Restoration approach using the 
Gmid
and (more extended) MBas basis sets (see Table~\ref{TBasises} for definition of these basis sets) as well as within the direct 4-component approach using the Amax basis set. In these calculations 18 valence electrons were correlated at the level of the coupled cluster with single amplitudes (CCS) as well as CCSD, CCSD(T) and CCSDT(Q) theories.
\begin{table}[h!]
\caption{Correlation contributions to \Eeff\ (in GV/cm) calculated within the 18-electron 2-component GRECP/Restoration (two-step) and direct 18-electron 4-component (4c) approaches using different basis sets.}
\label{HighCorr}
\begin{tabular}{lrrr}
\hline\hline
\backslashbox{Correlation\\contribution}{Basis set, \\ approach}     & 
Gmid 
& Mbas, & Amax, \\ 
    & two-step & two-step & 4c
\\
\hline                 
\Eeff(CCSD)-\Eeff(CCS)         & -9.9           & -11.9         & -11.2            \\
\Eeff(CCSD(T))-\Eeff(CCSD)     & -1.8            & -1.8          & -1.8             \\
\Eeff(CCSDT(Q))-\Eeff(CCSD(T)) &  0.0            & ---           & ---              \\
\hline\hline
\end{tabular}
\end{table}
It follows from Table \ref{HighCorr} that the correlation contributions to \Eeff\ from different cluster amplitudes calculated within the GRECP/Restoration and 4-component approaches almost coincide.
It justifies an application of the correction on the high order correlation effects to the considered parameters presented in Table \ref{TResults} within the GRECP/Restoration approach. The correction was estimated as the difference in the calculated parameters within the 18-electron CCSDT(Q) versus the CCSD(T) method.

One can also note a fast convergence of the coupled cluster series from Table \ref{HighCorr}. In Ref.~\cite{Skripnikov:15a} 
it was
also found that it is not necessary to use multireference coupled cluster approaches to the problem under consideration. Therefore, we have chosen single-reference CC approaches to calculate \Eeff\ and other AIC characteristics in the present paper.

\subsection{Performance of configuration interaction methods}
In Ref.~\cite{Skripnikov:15a} we found a poor convergence of the results of
calculation of AIC characteristics 
 within the single reference configuration interaction methods up to the configuration interaction with single, double, triple and quadruple excitations, CISDTQ. 
It was 
also noted that the value of \Eeff\ obtained within this most elaborate (among considered CI-approaches) method did not coincide with the final CC-based result. It was demonstrated using the CBas basis set within the scalar-relativistic GRECP/Restoration approach. Here we extend the treatment up to the configuration interaction with single, double, triple, quadruple and quintuple excitations, CISDTQP. To perform this calculation 
one
had to 
reduce further
the basis set size and construct the Gmin basis set (see Table~\ref{TBasises}).
The results for \Eeff\ are given in Table~\ref{TCISDTQP}.
By comparing them with that of Table 1 in Ref.~\cite{Skripnikov:15a}
%\footnote{Note that there are two misprints in Table 1 of Ref.~\cite{Skripnikov:15a}: for the CCSD(T) method and the CBas basis set correlation contributions to  \Eeff\ A$_{||}$ are 10.2 GV/cm and =-84$\frac{\mu_{\rm Th}}{\mu_{\rm N}}\cdot$MHz, respectively rather that the values (11.0 GV/cm for \Eeff and -103$\frac{\mu_{\rm Th}}{\mu_{\rm N}}\cdot$MHz for A$_{||}$) given there.} 
\cite{Note1}
 obtained in the larger CBas basis set one can note that the results of the Gmin basis set reproduce well the 	
ratios among different correlation methods.

\begin{table}[!h]
\centering
\caption{Correlation contributions to the effective electric field (\Eeff) and energy of the H$^3\Delta_1$ state of the ThO molecule in various 18-electron configuration interaction and coupled-cluster calculations relative to 2-electron CISD. Calculations were performed within the scalar-relativistic GRECP/Restoration approach using the minimal Gmin basis set.}
\label{TCISDTQP}
\begin{tabular}{lrrr}
\hline\hline
Method  & Correlation~~~      &  \Eeff (GV/cm) \\ 
        & energy (Hartree)~~~ &                \\ 
\hline
CCSD    & -0.247~~~             & 18.3                     \\
CCSD(T) & -0.258~~~             & 15.7                     \\
CCSDT   & -0.257~~~             & 15.7                     \\
CCSDTQ  & -0.258~~~             & 15.6                     \\
%        &                    &                          \\
\hline        
CISD    & -0.227~~~             & 16.5                     \\
CISDT   & -0.235~~~             & 18.3                     \\
CISDTQ  & -0.256~~~             & 16.5                     \\
CISDTQP & -0.257~~~             & 15.8                     \\
\hline\hline
\end{tabular}
\end{table}

One can see from Table~\ref{TCISDTQP} that CI- and CC- series do converge to the common value of \Eeff.
However, CC-series does it much 
quicker
-- even the CCSD(T) method gives almost the converged result (see also Table~\ref{HighCorr}). In the CI case one needs to consider up to quintuple excitations within the CISDTQP approach to get the converged result. Unfortunately, it is hardly possible to treat
to-date
 the ThO electronic structure within this method in a basis set with an adequate size even without the inclusion of the spin-orbit effects. One should note that the value of \Eeff\ calculated within the 4-component 18-electron multireference CI, MR$^{+T}_{12}$-CISD, approach (see Ref.~\cite{FleigComment:16} for the explanation of the abbreviation), 77.1 GV/cm, which was used in Ref.~\cite{FleigComment:16} as the corresponding base value of \Eeff\ differs from our value calculated within the 18-electron 4-component CCSD(T) method (see Table~\ref{TResults}) by -2.7 GV/cm (3\%). As the MR$^{+T}_{12}$-CISD method is an approximation to the CISDTQP method (which gives the value of \Eeff\ close to the one within the CCSD(T) method) we suggest that the main reason for the discrepancy is the lack of some types of important excitations in the MR$^{+T}_{12}$-CISD method. It covers only a certain subset from all  the excitations of the CISDTQP method. As is shown above even quintuple excitations can contribute nonnegligiably to the value of \Eeff.
A detailed analysis of important correlation contributions to \Eeff\ is also given in Ref.~\cite{Skripnikov:15a}.
Some of them are also missed in the MR$^{+T}_{12}$-CISD method 
%\footnote{Note that additional 18-electron corrections on ``$\Delta$ spinors'' (see \cite{FleigComment:16} for the definition) and the size of the active space calculated in Ref.~\cite{FleigComment:16} improves agreement with the value of \Eeff\ obtained within the CCSD(T) method. However, at the same time they decrease the agreement for the HFS constant.}
\cite{Note2}, 
e.g. all four-fold excitations from the closed-shell spinors, three-fold excitations from the closed-shell to the virtual spinors above the active space, etc.

\subsection{Inner-core and outer-core contributions}

The level of the accuracy considered in the present paper requires an estimation not only of the dominant contributions to the considered parameters from the outer-core and valence electrons but also smaller correlation contributions from the inner-core electrons of ThO. In the previous studies \cite{Skripnikov:15a,Skripnikov:13c,Fleig:14} these contributions were neglected. In Ref. \cite{Skripnikov:15a} 
it was
 mentioned that for the problem the size-extensive correlation methods, i.e. methods where the correlation energy properly scales with the number of correlated electrons should be used. Below we analyze it in more details and obtain a precise contribution to \Eeff\ from the inner-core electrons.

Table \ref{TMethodSelection} presents the results of calculations of outer-core correlation contributions to \Eeff\ within several methods to treat the electron correlation. The contributions were calculated as the difference in the calculated \Eeff\ values within the 38-electron versus the 18-electron coupled cluster and configuration interaction methods using the scalar-relativistic GRECP/Restoration approach. This approximation is enough to compare an applicability of the different correlation methods to the problem.
To apply such methods as the configuration interaction with single, double, triple and quadruple 
excitations, CISDTQ, with the treatment of 38 electrons we had to use a very small Gmin basis set.
By comparing the results calculated using this smallest basis set and using the basis sets of higher quality (CBas and MBas, see Table~\ref{TBasises}) one can see that the Gmin basis set reproduces roughly the main \textit{relative}  correlation contributions (which are given in parentheses) from the outer-core electrons, though it does not accurately reproduce the absolute values of the contributions, e.g. within the CISD method. 
%It seems however, that the Gmin basis set is enough for rough estimation of \textit{relative} correlation contributions.

\begin{table}[h!]
\centering
\caption{Contributions to \Eeff\ (in GV/cm) from the correlation of the outer-core electrons using different correlation methods and basis sets within the scalar-relativistic GRECP/Restoration approach. Relative values of \Eeff\ within a given series (e.g. in the CCS, CCSD, CCSD(T), CCSDT, CCSDTQ series or in the CIS, CISD, CISDT, CISDTQ series) with respect to the previous value in the series (e.g. \Eeff(CCSD)-\Eeff(CCS), \Eeff(CISDTQ)-\Eeff(CISDT), etc.) are given in parentheses.
The quality of the used basis sets increases in a line: Gmin, CBas and MBas.
}
\label{TMethodSelection}
\begin{tabular}{lrrr}
\hline\hline
%Method  & \Eeff, GV/cm & HFS \\ 
\backslashbox{Method}{Basis set}   & ~~~~~~~~~~Gmin & ~~~~~~~~~~CBas & ~~~~~~~~~~MBas \\ 
\hline
%\\
CCS     & 1.2 (---)~~~        &  2.7 (---)~~~  & 3.4 (---)~~~     \\
CCSD    & 1.6 ($+$0.4)  &  2.8 ($+$0.1) &  4.0 ($+$0.6)   \\
CCSD(T) & 1.1 ($-$0.5)  &  2.5 ($-$0.4) &  3.9 ($-$0.1)  \\
CCSDT   & 1.4 ($+$0.2)  &  2.6 ($+$0.1) &     \\
CCSDTQ  & 1.3 ($-$0.0)   &      &     \\
%        &              &     \\
\\
\hline        
CIS     &  1.2 (---)~~~       &  2.7  (---)~~~      &  3.4 (---)~~~     \\
CISD    & -1.5 ($-$2.7) &  0.5 ($-$2.3) &  0.1 ($-$3.3)   \\
CISDT   &  3.8 ($+$5.3) &  4.9 ($+$4.4) &     \\
CISDTQ  &  2.0 ($-$1.8) &      &     \\
\\
\hline 
MR3-CIS   &  1.2(---)~~~~ & 2.7 (---)~~~   &  3.4 (---)~~~    \\
MR3-CISD  & -1.4 ($-$2.6) & 0.6 ($-$2.2)   & 0.1 ($-$3.3)   \\
MR3-CISDT &  3.0 ($+$4.4) & 4.4 ($+$3.9) &     \\

\\
\hline
%        &              &     \\
QCISD   & 2.4 (---)~~~    & 2.7 (---)~~~ &  3.5 (---)~~~    \\
%QCISD(T) & ???     & 2.3 &  ???    \\
\hline\hline
\end{tabular}
\end{table}

One can see from Table \ref{TMethodSelection} that the 
coupled cluster 
approaches of different orders give very stable and close to each other results for the correlation contribution to  \Eeff\ from the outer-core electrons.
At the same time, not size-extensive configuration interaction methods do not demonstrate such a stability. 
Note that as 
one ``restores''
 the size extensivity by moving from the CISD method to the quadratic configuration interaction with single and double excitations, QCISD, method 
one obtains
the result close to that of 
the CCSD approach.

Table \ref{TMethodSelection} also includes
 the results of the multireference configuration interaction methods: within the MR(3)-CIS, MR(3)-CISD and MR(3)-CISDT methods. In these methods active spinor space includes six spin-orbitals corresponding to $7s_\sigma$ and 6d$_\delta$ orbitals of Th. The MR(3)-CIS method includes all possible single excitations from the closed shells and active spin-orbital space. The MR(3)-CISD method includes all possible single and double excitations from the correlated electrons while the MR(3)-CISDT method includes all possible single, double and triple excitations (note the difference between the MR(3)-CISDT and MR$_3$-CISDT methods used in Ref.~\cite{Fleig:14}, the latter does not include triple excitations from the closed-shell orbitals in opposite to the MR(3)-CISDT method).
The MR(3)-CISD method was used in Refs.~\cite{Fleig:14,FleigComment:16} by Fleig et al. to estimate the correlation contribution to \Eeff\ from the outer-core electrons. By comparing the values obtained within the MR(3)-CISD and MR(3)-CISDT methods (either in the Gmin or CBas basis sets) in Table \ref{TMethodSelection} one can see that the MR(3)-CISD method cannot be used to estimate accurately the correlation contributions from the outer-core electrons. This is due to the absence of the size-extensivity property of the method as well as the poor treatment of the electron correlation by this method as was demonstrated in Ref.~\cite{Skripnikov:15a}.

To calculate the correlation contributions from the inner-core electrons in the present study 
we have chosen coupled cluster
theory.

Table \ref{TShellsContribs} presents the calculated contributions to \Eeff\ from different 
groups of electrons
using the 
size-extensive methods within the 4-component approach as well as the GRECP/Restoration approach with spin-orbit effects included. The contributions from the correlation of the outer-core electrons were obtained as the difference in the calculated \Eeff\ values within the 36-electron (or 38-electron in the GRECP/Restoration case) versus the 18-electron approaches. 
Similarly, to extract the correlation contributions of the inner-core electrons we compared 98-electron and corresponding 36-electron calculations.

\begin{table}[!h]
\centering
\caption{Correlation contributions to \Eeff\ (in GV/cm) from the outer-core and inner-core shells using different methods within the direct 4-component (4c) and 2-component GRECP/Restoration approaches.}
\label{TShellsContribs}
\begin{tabular}{lrrrr}
\hline\hline
Shells (basis, approach)       & CCS & MP2(*)   & CCSD & CCSD(T) \\ 
\hline
Outer-core (Amin,4c) & 3.9 & 3.5      & 4.2  & 4.3     \\
Outer-core (Amid,4c) & 3.9 &       & 4.3   &     \\
Outer-core (Amax,4c)& 3.9 &   3.5   & 3.8  &4.2     \\
Outer-core          & 3.7  &  3.4        & 4.0  &4.3     \\
(MBas,GRECP/Restoration)&  &          &  &      \\
\hline
Inner-core (Amin,4c) & 3.3 & 2.9      & 3.6  & 3.3     \\
Inner-core (Amid,4c) & 3.3 &       & 3.6   &     \\
%\\
\hline\hline			
\end{tabular}
\\(*) MP2 is the M{\o}ller-Plessett perturbation theory of second order. The contribution was estimated as the first iteration of the CCSD calculation.
\end{table}

%It was not possible to compute contribution of inner-core electrons within the Amax basis set. Therefore, we used Amin basis set 

One can see from Table~\ref{TShellsContribs} that the contributions from the outer-core electrons calculated within the CC methods in the Amin and Amax basis sets almost coincide.
In addition, the contributions from the \textit{inner-core} electrons calculated using the Amin and Amid basis sets  also almost equal within the given (CCS of CCSD) method
%\footnote{As an additional test we have also estimated the inner-core contribution to \Eeff\ using the Amin2 basis set where $s-$, $p-$ and $d-$ Gaussians for Th were replaced by the alternative set of 26 $s-$, 23 $p-$ and 17 $d-$ Gaussians. They were constructed as follows. A number of numerical functions which include correlation functions for all shells of Th atom were generated. The functions were approximated by Gaussians within the even-tempered general-contracted scheme. The basis set was then used in an uncontracted form. The inner-core contribution to \Eeff\ was found to be very close (within 0.2 GV/cm) to the value obtained within the Amin basis set}. 
\cite{Note3}.
These two points suggest that estimation of the contribution from the correlation of the inner-core electrons within the Amin basis set is accurate enough. Besides, one can see that the leading correlation contribution from the inner-core electrons is already achieved at the CCS level. This suggests that the leading effect from the inner-core electrons is due to the spin-polarization of these electrons, though even triple cluster amplitudes do slightly contribute.
The correlation contributions to the considered parameters from the inner-core shells presented in Table~\ref{TResults} were estimated within the CCSD method using the Amin basis set.

It also follows from Table \ref{TShellsContribs} that the correlation contribution of the outer-core electrons calculated within the GRECP/Restoration approach reproduces well the direct 4-component one as was in the case of  the particular correlation contributions discussed above (see also Table~\ref{HighCorr}).

\subsection{Basis set correction}
The Amax basis set that was used in the main 4-component correlation calculation includes 5 $g-$ and 1 $h-$ type Gaussians (see Table \ref{TBasises}). It was not practically possible to use the basis set with considerably higher quality in the 4-component calculations within the same calculation parameters.
Therefore, we have applied the following two basis set corrections:
(i) Correction on extension of the number of $s-$, $p-$, $d-$ and $f-$ Gaussians on Th. For this corrections we have performed 4-component 18-electron CCSD calculations and considered the difference in the calculated parameters within the AmaxExt versus the Amax basis set.  
(ii) Correction on high angular momenta, i.e. $g-$, $h-$ and $i-$ type Gaussians on Th and on the extension of the basis set on oxygen. For this we have used the 38-electron scalar-relativistic CCSD(T) method within the GRECP/Restoration approach \textit{without} cutoff and considered difference in the calculated parameters within the GmaxExt2 versus the Gmax basis set. The latter includes 5 additional functions of $g-$ type, 9 of $h-$ type and 5 for $i-$ type with respect to the Gmax (Amax) basis set. 
%It was not practically possible to perform such calculations within the pure 4-component technique.
To justify the applicability of this correction within the scalar-relativistic GRECP/Restoration approach we selected two test basis functions (one of $h-$ type and one of $i-$ type Gaussians) and calculated their correlation contribution to the \Eeff\ and HFS parameters within the CCSD method with the treatment of the outer-core and valence electrons correlation.
In the 4-component case their contributions were found to be $-0.3$ GV/cm and $-3$ $\frac{\mu_{\rm Th}}{\mu_{\rm N}}\cdot$MHz to \Eeff\ and A$_{||}$, respectively.
In the 1-component GRECP/Restoration approach their contributions were found to be $-0.3$ GV/cm and $-1$ $\frac{\mu_{\rm Th}}{\mu_{\rm N}}\cdot$MHz to \Eeff\ and A$_{||}$, respectively, i.e. in a very good agreement.
%Summarized corrections on basis set extension are given in Table~\ref{TResults}.
Corrections (i) and (ii) on the basis set extension are given in Table~\ref{TResults} as ``Basis set correction 1'' and ``Basis set correction 2'', respectively.
%One can see that the total correction on basis set extension 
One can see from this Table that the basis functions with high angular momenta non-negligibly (-0.6 GV/cm) contribute to the value of \Eeff\ in spite of the fact that these functions cannot give considerable direct contribution as they have very small amplitudes in the vicinity of the Th nucleus. These basis functions also noticeably contribute to the transition energy and contributes about -0.04 D to the molecule frame dipole moment. Actually, it is not surprising for the compound of 5f actinide element thorium. 
Additional basis functions on oxygen in correction (ii) negligibly contribute to the values of \Eeff\ and A$_{||}$, but do contribute about +0.03 D to the value of molecule frame dipole moment in the considered $^3\Delta_1$ state of ThO.

\subsection{Gaunt contribution}
4-component CCSD(T) calculations presented in Table~\ref{TResults} were performed within the Dirac-Coulomb Hamiltonian. The contributions from the interelectron Breit interaction were estimated at the Hartree-Fock level as the difference in the values of the calculated parameters within the Dirac-Coulomb-Gaunt (in no-pair approximation) versus the Dirac-Coulomb Hamiltonian. 
In this treatment we used the version of the Hartree-Fock method which is optimal only for the $^3\Delta_1$ state in contrast to treatment within the average-of-configuration Hartree-Fock method used in all other calculations (see the COMPUTATIONAL DETAILS AND DEFINITIONS section).
The Gaunt contribution to \Eeff\ presented in Table~\ref{TResults} was obtained with the EDM operator given by Eq.~(\ref{Wd2}). In a similar way this correction to \Eeff\ was calculated in Ref.~\cite{FleigComment:16}.
For comparison with the results obtained within the EDM Hamiltonian (\ref{Wd2}) we have also estimated Gaunt contribution to \Eeff\ within the EDM Hamiltonian (\ref{Wd1}) in approximation when electric field $\bm{E}$ is produced only by the Th nucleus. The resulted contribution to \Eeff\ is -0.9 GV/cm. 
Note that in the Dirac-Coulomb-Gaunt approximation one should add some two-electron terms to both forms of the EDM operators in Eqs.~(\ref{Wd1}) and (\ref{Wd2}) \cite{Lindroth:89}. However, this is outside the scope of the present study. Therefore, the ``Gaunt contribution'' to \Eeff\ (-1.5 GV/cm) can be considered as an estimation of the uncertainly due to non-inclusion of the Gaunt interaction rather than the correction. Note also that in the GRECP/Restoration scheme one also takes into account some part of the Breit interaction within the GRECP approach \cite{Petrov:04b, Mosyagin:06amin}.

\subsection{Uncertainties}

The uncertainty on the correlation treatment is lower than 1\%. It follows from Table~\ref{TResults} by considering the correction on high order correlation effects. 
In addition it was found in Ref.~\cite{Skripnikov:15a} that the CCSD(T) method that was used here to compute the leading contributions to the considered properties is stable with respect to a choice of one-electron molecular spinors. This shows that the chosen method also treats accurately orbital relaxation effects. We 
have
also checked that the selected energy cutoff for 36-electron calculation, 50 a.u., is enough. For this 
\Eeff\ was calculated
at the 4c-CCSD(T) level  using Amin basis set with this cutoff and without any cutoff. The obtained values coincide within 0.005 GV/cm (0.006\%). Besides, we
have
 checked that the basis set correction is stable.
In Ref.~\cite{Skripnikov:15a} 
it was
 also showed that a particular choice of the nuclear model leads to changes in considered parameters lower than 1\%.
The main uncertainty of the current study is 
going 
from the approximate treatment of the Breit effects.
Taking into account the above analysis and the results on Table~\ref{TResults} we estimate the uncertainty of the final value of \Eeff\ to be lower than 4\%.

\section{Conclusion}

Our present study is the third one devoted to the electronic properties of ThO molecule. 
In the first 
treatment
 \cite{Skripnikov:13c} we considered the electronic structure of ThO within the two-step 2-component generalized relativistic effective potential followed by restoration of the 4-component electronic structure approach. The study included correlation treatment of the outer-core and valence electrons while the inner-core electrons were not considered. The uncertainty of the results was estimated as 15\%. In the second study \cite{Skripnikov:15a} we analyzed extensively different methods to treat the electron correlation effects and 
reported
the detailed correlation contributions to the considered properties. We have also investigated the high-order correlation corrections. The uncertainty of the treatment was estimated as 7\%.

To reduce further the uncertainty of the results in the present study we had to go beyond the pure GRECP/Restoration treatment. For this 
have
we developed the combined direct relativistic 4-component and two-step GRECP/Restoration scheme to  calculate precisely atoms-in-compounds characteristics such as the hyperfine structure constant, effective electric field and molecular constant of the scalar-pseudoscalar nuclear-electron interaction. 
The
code to compute the matrix elements of the corresponding operators over the multicenter four-component molecular spinors for linear molecules
was developed.
It was
shown that the GRECP/Restoration approach can reproduce accurately different correlation contributions to the AIC characteristics. Due to the flexibility of the GRECP/Restoration approach it was possible to consider corrections on the high-order correlation effects and
basis set extensions.

In the applied scheme we were able to include explicitly all 98 electrons (inner-core, outer-core and valence electrons) of the ThO molecule within the coupled cluster method, even within the single, double and noniterative triple cluster amplitudes, CCSD(T). This calculation included about $10^{12}$ cluster amplitudes.
For the valence electrons we were able to treat up to quadruple cluster amplitudes within the CCSDT(Q) method. 
The final value of \Eeff\ is close to the value obtained by us in Ref.~\cite{Skripnikov:15a}. This is partly due to  the cancellation of the uncertainty of the pure GRECP/Restoration approach used in the previous study and contributions from the new effects (correlation of the inner-core electrons)
first
 considered in the present study.
The obtained uncertainty of the parameters that are required to interpret the EDM experiment in terms of T,P-odd effects is estimated to be lower than 4\% and is 
almost twice
smaller than in the previous studies~\cite{Skripnikov:15a,Skripnikov:13c}.
At the present stage all the possible essential effects that can contribute to \Eeff\ are first considered and the reliability of the present study is dramatically improved compared to all previous ones.

Here we have 
 thoroughly extracted the correlation contributions 
also
 from the outer-core and inner-core electrons. In contrast to an assumption of other studies \cite{FleigComment:16} 
it was
shown that the contribution from the inner-core electrons is not negligible and contributes 4\% to \Eeff. The correlation of the outer-core electrons within the used size-extensive methods contributes 5\% to \Eeff. Thus, the summarized  correlation contribution to \Eeff\ from the outer-core and inner-core electrons achieves 9\%. This excides noticeably the estimation of the core-correlation contribution estimated as 1.2 GV/cm (1.5\%) in Ref.~\cite{FleigComment:16} and consequently the final uncertainty of 3\% of the calculated value of \Eeff\ in Ref.~\cite{FleigComment:16} seems to be
notably
 underestimated. To explain this 
it is demonstrated
that the multireference configuration interaction method (MR3-CISD) that was used in Ref.~\cite{FleigComment:16} to consider the contribution to \Eeff\ from the outer-core electrons correlation cannot be applied for an accurate extraction of the contribution. As one includes higher order correlations, e.g. within the MR3-CISDT method the core contribution changes dramatically with respect to the MR3-CISD estimation.

The developed calculation scheme and code can be used to consider
most
 precisely other promising systems such as ThF$^+$ cation, TaN molecule, etc. to the search for \eEDM\ and other T,P-odd effects in heavy-atom molecules and atoms.

\section{Acknowledgement}
I am grateful to Anatoly Titov for valuable discussions and remarks.
Molecular calculations were partly performed on the Supercomputer ``Lomonosov''.
The development of the code for the computation of the matrix elements of the considered operators as well as  the performance of all-electron calculations were funded by RFBR, according to the research project No.~16-32-60013 mol\_a\_dk. GRECP/Restoration calculations were performed with the support of President of the Russian Federation Grant No.~MK-7631.2016.2 and Dmitry Zimin ``Dynasty'' Foundation.

%\bibliographystyle{./bib/apsrev}
%\bibliography{bib/JournAbbr,bib/SkripnikovLib,bib/QCPNPI,bib/TitovLib,bib/Kaldor,bib/PetrovLib,bib/Lomachuk,bib/Kudashov}

\end{document}